\documentstyle[12pt]{article}

\begin{document}

\topmargin 0pt
\oddsidemargin 5mm
\def\bbox{{\,\lower0.9pt\vbox{\hrule \hbox{\vrule height 0.2 cm
\hskip 0.2 cm \vrule height 0.2 cm}\hrule}\,}}
\def\a{\alpha}
\def\b{\beta}
\def\g{\gamma}
\def\G{\Gamma}
\def\d{\delta}
\def\D{\Delta}
\def\e{\epsilon}
\def\h{\hbar}
\def\ve{\varepsilon}
\def\z{\zeta}
\def\t{\theta}
\def\vt{\vartheta}
\def\r{\rho}
\def\vr{\varrho}
\def\k{\kappa}
\def\l{\lambda}
\def\L{\Lambda}
\def\m{\mu}
\def\n{\nu}
\def\o{\omega}
\def\O{\Omega}
\def\s{\sigma}
\def\vs{\varsigma}
\def\S{\Sigma}
\def\vphi{\varphi}
\def\av#1{\langle#1\rangle}
\def\pa{\partial}
\def\na{\nabla}
\def\hg{\hat g}
\def\un{\underline}
\def\ov{\overline}
\def\cF{{{\cal F}_2}}
\def\Hsl{H \hskip-8pt /}
\def\Fsl{F \hskip-6pt /}
\def\cFsl{\cF \hskip-5pt /}
\def\ksl{k \hskip-6pt /}
\def\pasl{\pa \hskip-6pt /}
\def\tr{{\rm tr}}
\def\tcF{{\tilde{{\cal F}_2}}}
\def\tg{{\tilde g}}
\def\shalf{\frac{1}{2}}
\def\nn{\nonumber \\}
\def\w{\wedge}

\def\cmp#1{{\it Comm. Math. Phys.} {\bf #1}}
\def\cqg#1{{\it Class. Quantum Grav.} {\bf #1}}
\def\pl#1{{\it Phys. Lett.} {\bf B#1}}
\def\prl#1{{\it Phys. Rev. Lett.} {\bf #1}}
\def\prd#1{{\it Phys. Rev.} {\bf D#1}}
\def\prr#1{{\it Phys. Rev.} {\bf #1}}
\def\prb#1{{\it Phys. Rev.} {\bf B#1}}
\def\np#1{{\it Nucl. Phys.} {\bf B#1}}
\def\ncim#1{{\it Nuovo Cimento} {\bf #1}}
\def\jmp#1{{\it J. Math. Phys.} {\bf #1}}
\def\aam#1{{\it Adv. Appl. Math.} {\bf #1}}
\def\mpl#1{{\it Mod. Phys. Lett.} {\bf A#1}}
\def\ijmp#1{{\it Int. J. Mod. Phys.} {\bf A#1}}
\def\prep#1{{\it Phys. Rep.} {\bf #1C}}

\begin{titlepage}
\setcounter{page}{0}

\begin{flushright}
IASSNS-HEP-98/5 \\
hep-th/9801162 \\
January 1998
\end{flushright}

\vspace{5 mm}
\begin{center}
{\large Radiation of RR states from NS Fivebranes }
\vspace{10 mm}

{\large S. P. de Alwis\footnote{e-mail: dealwis@sns.ias.edu, 
dealwis@gopika.colorado.edu}~\footnote{On leave from 
Department of Physics, Box 390,
University of Colorado, Boulder, CO 80309. }}\\
{\em School of Natuaral Sciences, Institute for Advanced Study, 
Princeton NJ 08540}\\
\vspace{5 mm}
\end{center}
\vspace{10 mm}

\centerline{{\bf{Abstract}}}
We show that the recent observation that near extremal NS fivebranes
decay due to Hawking radiation, can be understood in terms of their coupling
to  RR states which does not vanish even as the string coupling goes to
zero. 

\end{titlepage}
\newpage
\renewcommand{\thefootnote}{\arabic{footnote}}
\setcounter{footnote}{0}

\setcounter{equation}{0}
In a recent paper \cite{ms} it has been argued that systems of
 NS fivebranes in type IIA and type IIB string theory, 
which are expected to decouple from the bulk ten dimensional theory in
the limit when the string coupling goes to zero \cite{ns},
in fact emit Hawking radiation into the bulk\footnote{There is a hint
of this phenomenon in the works cited in \cite{ss}.}. 
In this note the source of this radiation
is identified. It is argued that while the NSNS states of the bulk
theory do decouple in the relevant limit, the RR states do not, provided
we work self-consistently by using the induced metric from the
bulk  on the brane. On the other hand if we take the flat metric 
on the brane then it decouples from all bulk modes. The latter is 
perhaps to be identified with the decoupled theories of \cite{ns}
but they should not be considered to be the source of the supergravity
fivebrane. The latter needs to be treated self-consistently.
 We also take the matrix model limit 
 \cite{jm2},\cite{as},\cite{ns2} and show that 
 the system of NS fivebranes that is thus obtained, does infact decouple 
 and there is no Hawking radiation into the bulk supergravity 
 coming from string theory. However there is  radiation
 into the space defined by classical configurations of the moduli space 
 of the matrix model. 

The bosonic part of the effective theory on the M-theory fivebrane 
is given by the following Lagrangian.
\begin{eqnarray}\label{fiveaction}
I &=&{1\over (2\pi)^5l_p^6} \big [\int_{W_6}d^6\xi 
\sqrt{\det g_{\mu\nu}\pa_aX^{\mu}\pa_bX^{\nu}} \nn
& &+{1\over 4}\int_{W_6} (db_2-P(C_3)\wedge *(db_2-P(C_3)) +
{1\over 2.3!}\int_{W_6} dP(C_3)\wedge b_2
\big ] 
 \end{eqnarray}

In the above $b_2$ is the chiral two form living on the world volume of the 
five brane and strictly speaking there is no action for it\footnote{See the 
discussion in \cite{ew}.}. We are however
using the above only to determine the couplings to bulk (weak) 
supergravity fields.
So for our purposes it is sufficient to impose the self duality constraint
at the level of the equations of motion. $C_3$ is the 
three form field of 11-D supergravity and $P()$ is the pullback
map from the eleven manifold to the world volume $W_6$ of the fivebrane. Its
presence is required by the fact that M2-branes can end on these 5-branes
\cite{as}\cite{pt}.
\footnote{It should be
noted that although the above action is not gauge invariant by itself under the
gauge variation $\d C_3 = d\L, \d b_2=\L$ this `anomaly' is cancelled by
anomaly inflow from the bulk supergravity action\cite{ew}\cite{sda}. For the
complete action including the coupling of the M2-branes and references to
earlier work see \cite{sda}.}

The NS fivebrane of type IIA string theory is obtained from the M-theory
fivebrane when the circle on which M-theory is  compactified is taken to
be transverse to the fivebrane.\footnote{If this circle is in 
a longitudinal direction then we obtain the D4-brane of type IIA.}
The relation between the metrics and the anti-symmetric tensor fields of the
two supergravities are given by,
 \begin{eqnarray}\label{metric}
 ds^2_M&=&e^{-2\phi/3}ds^2_{IIA}+e^{4\phi/3}(dy-C_1)^2.  \nn
C^{(M)}_3& & C^{(IIA)}_3+3B_2\w dy.
\end{eqnarray}
So we have for the IIA fivebrane
\begin{eqnarray}\label{fiveaction}
I &=&{1\over (2\pi)^5l_s^6} \big [\int_{W_6}d^6\xi e^{-2\phi}
\sqrt{\det g_{\mu\nu}\pa_aX^{\mu}\pa_bX^{\nu}} \nn
& &+{1\over 4}\int_{W_6} (db_2-P(C_3)\wedge *(db_2-P(C_3)) +
{1\over 2.3!}\int_{W_6} dP(C_3)\wedge b_2
\big ] 
 \end{eqnarray}
 It should be noted that in the above (as well as in all equations below)
 $l_s=l_p$ is the string scale determined in the string metric while
 $l_p$ is the M-theory Planck scale determined in the M-metric.
 
The bulk low energy supergravity action of IIA string theory has the 
action 
\begin{eqnarray}\label{twoa}
I_{IIA}&=&-{1\over 2\k^2}\int_{M_{10}}\sqrt{-G}e^{-2\phi}\left\{ R-4(\nabla\phi
)^2+{1\over12}H^2 \right\}\nn
& &-{1\over 2\k^2}\int_{M_{10}}(\shalf F_2\w *F_2+\shalf F_4\w *F_4)-{1\over
2\k^2}\shalf\int_{M_{10}}F'_4\w F'_4\w B_2,
\end{eqnarray}
where
\begin{equation}\label{mtwoa}
2\k^2=(2\pi)^7l_s^8,~ H_3=dB_2,~F_2=dC_1,~F'_4=dC_3,~
F_4=F'_4-H_3\w C_1.
\end{equation}
Corresponding to the fivebrane whose effective action was written down above
there is a solution of the equations of motion of this supergravity \cite{hs}.
 i.e.
\begin{eqnarray}\label{background}
 ds^2&=& -\tanh^2\s dt^2+(g^2l_s^2\mu\cosh^2\s +kl_s^2)(d\s^2+d\O_3^2)
 +dy_5^2 \nn
e^{2\phi}&=&g^2+{k\over\mu\cosh^2\s}  \nn
H&=& k\e_3l_s
\end{eqnarray}
In the above the coordinates $y_5$ represent the directions along the
fivebrane and $\e_3$ is the volume element on the unit three sphere.
 The  coordinate $\s$ parametrizes only
the region outside the horizon which is at $\s =0$.
The usual Schwarzchild coordinate is related to $\s$ by
the equation  
\begin{equation}\label{r}
r=r_0\cosh\s, ~~r_0^2=g^2\mu l_s^2.
\end{equation}
Note that in these coordinates the horizon is at $r=r_0$.
The ADM mass of the five brane is 
\begin{equation}\label{adm}
M={V_5\over (2\pi )^5 l_s^6}({k\over g^2}+\mu ),
\end{equation}
and we may identify $\mu$ as the dimensionless energy density on the
brane \cite{ms}.

We identify the source of the Hawking radiation by following the procedure
of Das and Mathur \cite{dm}
  To get canonically normalized kinetic terms for fluctuations around these
background fields we need to put for the NSNS sector,
\begin{equation}\label{}
g_{\mu\nu}=g_{\mu\nu}^{(0)}+g\k h_{\mu\nu},~~B_{\mu\nu}=B_{\mu\nu}^{(0)}+
g\k b^{(1)}_{\mu\nu},~~\phi=\phi^{(0)}+g\k\varphi .
\end{equation}
where the superscript $(0)$ refers to the background solution given by
(\ref{background}). It should be noted here that the relevant normalization
is that of the asymptotic fields (i.e. for $\s\rightarrow\infty$) where
the effective $e^{2\phi}=g^2$. This is what would go into an S-matrix
calculation of the emission process as in \cite{dm}.

For the RR fields however the properly normalized\footnote{It should be
remembered that our definition of $\k$ does not involve factors of $g$.} 
fluctuations are 
(the background RR fields are zero for this NSNS brane)
\begin{equation}\label{}
C_1=\k c_1~~C_3=\k c_3.
\end{equation}
 Note that in this limit the background
dilaton term (at the horizon) that goes into the first term of
\ref{fiveaction} is $e^{-2\phi(0)}={\mu\over k})$ though
we should actually put the 
brane at a `streched horizon' where $\s \sim O(1)$ since there is
a coordinate singularity at $\s =0$.  

These scalings show us that in the $g\rightarrow 0$ limit with the
energy density $\mu$ on the brane fixed, NSNS 
fluctuations will decouple.
However 
the RR fluctuations do not decouple and one has the following 
suviving couplings.
\begin{equation}\label{}
\Delta I ={1\over (2\pi )^5l_s^6}[{1\over 4}\int_{W_6} 
(db_2-P(C_3)\wedge *(db_2-P(C_3)) +
{1\over 2.3!}\int_{W_6} dP(C_3)\wedge b_2
\end{equation}
In the above the pull back of $C_3$ is explicitly,
\begin{equation}\label{}
P(C_3)_{[bcd]} = C_{\nu\l\s}\pa_bX^{\nu}\pa_cX^{\l}\pa_dX^{\s}.
\end{equation}
Thus we see that excitations of the moduli fields $X$ of fivebrane
 can annihilate
to produce Hawking radiation of the $C_3$ field.

A comment on the fact that by itself the 
(second integral of  the) five-brane
Lagrangian is not gauge invariant under $\d b_2=\L_2,\d C_3=d\L_2$ is in order
here. As in the 
M-theory case the problem stems from the fact that the five-brane
couples naturally to the six-form gauge field $C_6$ whose field
strength is dual to that of $C_3$. However the bulk supergravity
action contains only the $C_3$ form field (and there is no dual 
formulation of the  supergravity action). In the M-theory case \cite{sda}
the resolution involved introducing a particular solution to the modified 
Bianchi identity. The present case is simply the dimensionally 
reduced one coming from that argument. Thus we solve
 $dH_3=2\k^2\d (M_{10}\rightarrow W_6)$ in the presence of
 the five-brane by writing $H_3=H_3'+\t_3$ where $dH_3'=0,~d\t=\d,~ d*\t=0$.
 The Chern-Simons like term in the bulk action (\ref{twoa}) 
 is then modified as follows 
 \begin{equation}\label{}
-{1\over
2\k^2}\int_{M_{10}}[dC_3\w dC_3\w B_2+\shalf C_3\wedge dC_3
\wedge (dB_2+\t_3)].
\end{equation}
It can be seen explicitly that the gauge variation of this bulk term 
cancels the variation of the world volume action.

The invariance under the transformations $\d b_2~\L_2, etc$ implies that 
the $b_2$ field can be gauged away. Thus we conclude that in such a 
gauge the  entire
coupling of the five-brane to the RR field of the bulk comes from
a term of the form 
\begin{equation}\label{}
\int P(C_3)\w *P(C_3) .
\end{equation}

Similar arguments can be made for the (1,1) gauge theory on the
type IIB NS five brane. In this case there is for instance
a  coupling of the four form $C_4$ whose field strength is self-dual. This
may be obtained by S-duality from the corresponding D5-brane action, namely
\begin{equation}\label{}
{1\over (2\pi)^5l_s^6}\int (F_2-P(C_2))\w P(C_4).
\end{equation}
(Note that the field $C_2$ is the dual of $B_2$ and $C_4$ is self dual under
$SL(2,Z).$)
 
The relevant bulk term is
\begin{equation}\label{}
{1\over 2\k^2}\int_{M_{10}}\left (\shalf F_3\w *F_3+\shalf F_5\w *F_5+
C_4\w H_3\w F_3 \right ) ).
\end{equation} 
The comments about actions for 
self-dual fields made after (\ref{fiveaction}) apply in the above expression
too. Also as in the
IIA case an anomaly under the gauge transformation $\d C_4=d\L_3$ is
cancelled by anomaly inflow from the bulk,
and the rescaling to  get properly normalized
bulk fields does not involve factors of $g$, so that the coupling of these
bulk RR fields to the NS fivebrane survives the limit $g\rightarrow 0$.

 Let us now verify that these arguments are
consistent with the observations of \cite{ms},\cite{gs},
on the supergravity side. The Hawking temperature is identified in these
papers as the period (in Euclidean time) of the  Euclidean section
of the five-brane metric. In the $g\rightarrow 0$ case this gave
\begin{equation}\label{temp}
T_H ={1\over 2\pi l_s\sqrt k}.
\end{equation}
This result appears to be rather puzzling at first sight. The Hawking 
temperature is a semi-classical effect and should vanish in the limit
that the effective $\hbar$ i.e. the loop counting parameter is zero.
In string theory (in natural units) this parameter is the asymptotic
value of the dilaton namely $g^2$ and one should expect the temperature
to vanish when this parameter is zero. To clarify the issue let us
review the usual argument. 

Consider the propagation of quantum fields with a Lagrangian 
$L(\phi,\pa\phi)$
in some non-trivial  background metric. The partition function may be 
represented as a Euclidean functional integral with periodic boundary
conditions; i.e. we may write
\begin{equation}\label{}
\tr e^{-\b H}=\int d\phi e^{-{1\over\hbar}\int_0^{\Theta_H}L}.
\end{equation}
Here $\Theta_H$ is the period with which the Euclidean time is identified
in order to avoid a conical singularity at the origin (which is the
horizon). This enables us to identify the Hawking temperature as 
\begin{equation}\label{}
T_H=\b^{-1}={\hbar\over \Theta_H}.
\end{equation}
 
 Now in IIA string theory the effective action (for NSNS states)
  is given by the terms in the first integral in (\ref{twoa}) and these
  have an explicit factor of $e^{-2\phi}$ and hence one should take the
  effective $\hbar$ to be $g^2$ and therefore
 the temperature should go to zero for the emission of
 these states. However  the RR field couplings are given by the second 
 integral which does not have a factor of the dilaton and so 
 the effective $\hbar$ is one. This explains why the correct result
 for the temperature is obtained by just calculating the periodicity
 of the Euclidean section. Similar arguments can clearly be made for type
 IIB as well.
 
 It should be stressed that in the above discussion we have used the
 metric induced from the bulk as the metric on the brane. This is the 
 consistent choice if we want the brane to be the source of the corresponding
 supergravity metric. i.e. this corresponds to considering the
 coupled action $S[g]+I[g]$ where the two actions are the supergravity
 and brane actions, and taking dervatives with respect to $g$ etc to get
 equations with a ssource term. On the other hand in much of the
  literature on
 brane dynamics leading to Hawking radiation etc. uses the flat metric
 on the brane. However if one were to do this in this case it can be 
 immedeately seen that all modes will decouple\footnote{I wish to 
 thank J. Maldacena for this comment.}. For in this case the kinetic 
 term for the moduli looks like
   $g^{-2}\int \sqrt{\det \eta_{\mu\nu}\pa_aX^{\mu}\pa_bX^{\nu}}$. Then
    we need to rescale the moduli by writing
   $U=X/g^{2\over 3}$ and hold $U$ fixed as we take the limit 
   $g\rightarrow 0$. Now it is clear that all couplings to the bulk
   disappear; for instance the RR coupling $\int (c_3(\pa X)^3)^2$ goes
   to zero like $g^4$. This brane clearly cannot radiate and presumably
   should be identified with the isolated NS 5-brane theory of Seiberg\cite{ns}.
   On the other hand if we wish to make a connection to the supergravity
   solution corresponding to the fivebrane then we should be looking at
   a self-consistent solution as discussed earlier.

Let us now discuss what happens in the matrix
model limit \cite{jm2},\cite{as},\cite{ns2}.
In this case the limit to be taken is not  $g\rightarrow 0$ with $l_s$ fixed,
but 
\begin{equation}\label{xm}
g,l_s\rightarrow 0;~~ {g\over l_s^3} =g^2_{YM}\equiv l_m^{-3},
x_m={l_m^2\over l_s^2}x =l_m^2U fixed.
\end{equation}
 Where $x$ is any spatial coordinate. The last condition is just the statement
 that the gauge field is kept fixed as one goes to the  limit. $l_m$ is 
 effectively the Planck length of the supergravity that is generated by the 
 matrix model. Since we are
 rescaling spatial coordinates we should also rescale the time 
 (in order to keep the velocity of light fixed at $c=1$) and we write
  \begin{equation}\label{tm}
t_m={l_m^2\over l_s^2}t.
\end{equation}
In the above as well as below the subscript $m$ will denote quantities
that have been rescaled in accordance with the matrix model requirements.
By compactifying on a five torus and T-dualizing we get an effective string
coupling  
The ADM mass per unit volume in the string coordinates is given in
(\ref{adm}). In the limit considered in \cite{ms} (i.e. $g\rightarrow 0, ~
l_s, \mu$ fixed) the extremality parameter $r_0\rightarrow 0$.
In the matrix model limit the appropriate mass per unit
volume is 
\begin{equation}\label{}
{M_m\over V_{5m}}={l_s^{12}\over l_m^{12}}{M\over V_5}={k\over (2\pi)^5l_m^6}
+{l_s^4\over (2\pi)^5l_m^{12}}\mu.
\end{equation}
Note that in terms of the rescaled extremality parameter 
\begin{equation}\label{}
\mu={l_m^2r_{0m}^4\over l_s^4}.
\end{equation}
Thus taking the  limit with $\mu$ fixed  implies that the rescaled 
extremality parameter goes to zero. Thus the physical situation is the analog 
of that considered in \cite{ms} but with the rescaled variables. However
now the expression for the temperature gets rescaled as follows (note that this
is simply the inverse of the rescaling of the time).
\begin{equation}\label{}
T_m={l_s^2\over l_m^2}T={l_s\over 2\pi\sqrt{k}l_m^2}.
\end{equation}
This goes to zero as $l_s$ goes to zero. It is easy to check that this is
consistent with the microscopic picture. The RR coupling to the five brane is
(schematically)
\begin{equation}\label{}
{1\over l_s^6}
\int (C_{...}(\pa X)^{3...})^2=
{l_s^8\times l_s^{12}\over l_s^6l_m^{12}}\int (c_{...}(\pa X_m)^{3...})^2 .
\end{equation}
This goes to zero in the limit.

The above discussion is of course for (transverse) five branes which are
to be regarded as composites of zero-branes. Unlike the case of the membrane
and the longitudinal fivebrane there is no clear evidence that this in fact
exists. On the other hand we may compactify on a five torus and take the matrix
model limit. Using S and T duality it has been shown in \cite{jm2},\cite{as},
\cite{ns2} that this gives the (limit of) the NS fivebrane in type IIB. 
The decoupling argument will follow on the same lines as above.

Thus we see that the matrix model limit gives decoupled five-brane theories
which do not radiate even into the infinite tube and are
strictly six dimensional.This rather
than the original limit ($g\rightarrow 0,~l_s$ fixed) appears to be
 the proper limit to get decoupled theories.
 
 One final comment is that although the matrix model is decoupled 
 from the original bulk string theory it is supposed to regenerate
 the supergravity metric that we are studying as a classical 
 configuration of its moduli space. In this space one should then see
 the Hawking radiation etc that was suppressed in the original (string)
 theory. It is easily seen that the corresponding situation is obtained
 by using (\ref{xm}), (\ref{tm}) to rewrite the metric as
 
\begin{eqnarray}\label{}
ds_m^2 &\equiv&{l_m^4\over l_s^4}ds^2=-(1-{r^2_{m0}\over r^2_{m}})dt_m^2  \nn
&+&(1+{k_ml_m^2\over r_m^2})({dr_m^2\over 1-{r_{m0}^2\over r_m^2}}+r_m^2d\O^2_3)
+dy_{m6}^2
\end{eqnarray}
where we have defined $y_m={l^2_m\over l^2_s}y$ and ${k_m\over l_m^2}=
{k\over l_s^2}$. Effectively now the original string length (which 
goes to zero) is now replaced by $l_m$ which remains finite in the limit.
Thus there is now a finite temperature (\ref{temp}) and a corresponding 
non-vanishing microscopic coupling to the bulk fields
 which are now the bulk fields of
the regenerated supergravity coming from the matrix model. All this of course
assumes that the matrix model moduli space correctly
reproduces supergravity.

{\bf Acknowledgments:} I'm grateful to  Samir Mathur for several valuable
discussions. I would also
like to thank Micha Berkooz, Steve Giddings, Gary Horowitz, 
Rob Leigh, J. Maldacena, Rob Myers, Nathan Seiberg and Savdeep Sethi 
for  useful discussions,
 and Edward Witten for hospitality at the Institute for
Advanced Study and  the Council on  Research and Creative Work
 of the University of Colorado
for the award of a Faculty Fellowship. This work is partially supported by
the Department of Energy contract No. DE-FG02-91-ER-40672.


\end{document}